# The Multiverse Origin of our Physics does without Strings, Big Bang, Inflation, or Parallel Universes

Tom Gehrels

*Department of Planetary Sciences, University of Arizona*

December 29, 2009

Evolution needs long times and large numbers of samples or species. Our finely tuned physics can therefore not have evolved during the fast changes of a single Big-Bang universe, but the scales of $10^{30}$ years and $10^{19}$ universes of our multiverse satisfy that condition. Planck and Chandrasekhar equations show that multiverse, and a variety of observations show the origin of our physics.

Our multiverse is being fed by the debris of its decaying universes, it is transported on accelerated expansion. New universes originate from clouds of that debris, which is re-energized by the gravity at the cloud-centers when the proton density of $10^{18}$ kg/m$^3$ is reached. That epoch occurs ~$10^{-6}$ seconds, *i.e.* $10^{37}$ Planck times, later than a Big Bang. It marks the beginning of our universe with a photon burst that may have been observed by WMAP as the radiation signature with a wider curvature than that of the 3-K radiation. Karl Schwarzschild's test also confirms that beginning.

Key words: physics, protons, universes, dark energy, physical constants.

## CONTENTS

## 1. INTRODUCTION

This paper is a reply to the concise discussions of how our physics has gone awry with strings and interpretation of quantum theory by Smolin (2007) and by Penrose (in Kruglinski 2009). Previous titles are in www.lpl.arizona.edu/faculty/gehrels.html while now I try their concise approach for greater effect.

There are two distinct derivations, the first for the multiverse, which is based without anthropic assumptions on the equations of Max Planck (1858-1947) and Subrahmanyan Chandrasekhar (1910-1995). Planck defined units for mass, time, length and temperature in terms of the Planck constant h, radiation velocity c, gravity constant G, and Boltzman constant k (Planck 1899). Chandra derived an expression for cosmic masses in terms of h, c, and G, and of proton mass H (Chandrasekhar 1951). The paper begins where Chandra left off, by comparing his expression with the Planck domain. When this leads to a multiverse, a verification is inserted, but then its results are encouraging to proceed (Sec. 4). The surprise is that the multiverse has a precisely known mass, physics, and evolution for its universes, all being the same as in our universe.

The second part of this paper answers the question, "how did our universe originate from the multiverse?" This part is simpler than the first, because it is based on dozens of observations for our universe's aging and debris, and for the early stages of our universe. A theoretical confirmation is provided by a 1916 paper of Karl Schwarzschild (1873-1916).

The outline of this paper is in the above Contents; it ends with a summary of 37 conclusions and suggestions for future work.

## 2. PLANCK AND CHANDRASEKHAR EQUATIONS

Planck derived the theory of blackbody radiation and found units with them for length, mass, time and temperature through dimensional analysis; for instance the Planck mass is $(hc/G)^{0.5}$. Modern values of the constants are h = 6.626 0693(11) x $10^{-34}$ $m^2$ kg $s^{-1}$, c = 299 792 458 m $s^{-1}$ (in a vacuum, exact by definition), while G = 6.6742(10) x $10^{-11}$ $m^3$ $kg^{-1}$ $s^{-2}$ (Mohr & Taylor 2005). The numbers in parentheses are the estimated standard deviations; the relative standard uncertainty of G, for example, is 1.5 x $10^{-4}$. The proton mass is 1.672 621 71(29) x $10^{-27}$ kg.

Chandra developed the theory of structure, composition, and source of energy for stars involving a variety of physical laws such as of Stefan and Boltzmann, which relates pressure and temperature at various depths inside the star (Chandrasekhar 1951, pp. 599-605). The constants h, c, G, and H appear in the physical laws of the stars - which are the basic components of the cosmos - and they represent aspects of quantum, relativity, gravity, and atomic theory in unified operation. For the total stellar mass he thereby derived,

$$M = (hc/G)^{1.5} H^{-2}. \qquad (1)$$

He had also discovered a generalization for cosmic masses,

$$M(\alpha) = (hc/G)^{\alpha} H^{1-2\alpha}, \qquad (2)$$

for positive exponents, which identify the type of object, such as with $\alpha$ = 2.00 for our universe, in addition to the above $\alpha$ = 1.50 for stars.

## 3. THE "UNIVERSAL PLANCK MASS", M(α)

### 3.1. Restricted to Primordial Baryonic Objects

The paper deals primarily with baryonic and primordial masses. In the case of stars, the usage of M($\alpha$) is limited to original matter consisting mostly of hydrogen and helium in O and B stars, from which formed the compositions in subsequent stars with increased abundances of heavier elements.

### 3.2. Calibrated with the Proton Mass

A simplification of Eq. (2) is possible by expressing the masses in terms of the *universal mass unit of the proton mass*, such that H = 1, and

$$M(\alpha) = (hc/G)^\alpha, \tag{3}$$

provided that all masses be expressed in terms of the proton mass, H. One could call this expression the "universal Planck mass", considering it "… one of the most beautiful and important formulae in all of theoretical astrophysics…" (Shu 1982).

### 3.3. Overview Table

Table 1 presents data in the universal unit of the proton mass. These are representative proton masses; it does not say that there are $10^{78}$ protons in our universe. Comparison with observations was made already by Chandra in the 1930s, namely $1.1 \times 10^{78}$ proton masses for the universe that had been obtained from star counts, and in his Nobel-prize lecture he has 29.2 solar masses for O and B stars [$3.469\ 96(79) \times 10^{58}$ proton masses; see Stahler et al. 2000]. The preparations were pursued in Gehrels (2007a,b, 2009); a search for other objects participating in $M(\alpha)$ was made and a value of $\alpha = 1.00$ found for planetesimals, but this needs further study.

**Table 1. Masses computed for various Objects**

| $\alpha$ | Proton masses | Type of Object |
|---|---|---|
| ↑ | | |
| 3.00 | $1.2 \times 10^{117}$ | Cluster of Universes |
| 2.50 | $3.7 \times 10^{97}$ | Local Group of Universes |
| 2.00 | $1.1 \times 10^{78}$ | Universe |
| 1.50 | $3.5 \times 10^{58}$ | O and B Stars |
| 0.50 | $3.3 \times 10^{19}$ | Planck mass |
| 0.00 | 1 | Proton |

### 4. THREE CONSTANTS LOCK THE PROTON RADIUS

This section is to gain confidence in some of the concepts such as the Planck mass and the finite mass of our universe; it is in three stages. The Planck mass had been theorized in an original phase of matter, at an impossibly high Planck density, compressed within one cubic Planck length such that most of its components, but not all, can interact at velocity c, which is a Planck length in a Planck time. These concepts define the Planck mass, while we now are replacing that with its role in the mass scaling of the cosmos (Table 1). We also need $c^5/hG^2$, which is the computed

$$\text{Planck density} = 8.2044(25) \times 10^{95} \text{ kg m}^{-3}. \tag{4}$$

The constant factor F between steps of $\Delta\alpha = 0.50$ in the Table is seen in the number of proton masses for the Planck mass, which is the same number as for primordial stars generated in our universe,

$$F = 3.261\ 68(25) \times 10^{19}. \tag{5}$$

The ratio of the universe's mass [$1.131\ 79(35) \times 10^{78}$ proton masses] and the Planck mass [$3.261\ 68(25) \times 10^{19}$] has the third power of F, but the third root of that is taken for the length ratio from that volume ratio, coming back to F. Thus we obtain a size parameter for the universe at Planck density from the product of F and Planck length. That is however the size of a rib of a cube, while for the radius of a spherical volume for the universe one divides by the cube root of $4\pi/3$ to obtain R' = $8.1974(9) \times 10^{-16}$ m.

A basic exercise but without demonstration of the cube in the explanation of the Planck mass, is to divide the mass of the universe in Table 1 by the Planck density and obtain radius R' again.

A more precise derivation is made by realizing that the *formulae* for the universe's mass in Eq. (2), $(hc/G)^2 H^{-3}$ [H also in kg], divided by the one for Planck density, $c^5/hG^2$, yield the volume of $h^3 c^{-3} H^{-3}$. The low-precision gravity term, G, now takes no longer part and that increases the precision of the derivation. After rib-radius conversion again, the radius of the universe, if it would ever have been at the unlikely Planck density, would have been,

$$R = 8.197\ 3725(20) \times 10^{-16}\ m, \qquad (6)$$

with the precision depending only on those of h and H, since c is exact and G is no longer involved. An anonymous referee commented that this derivation is in relativistic theory, which is a most interesting remark but I have not been able to verify it, and the Editor did not allow reaction from the author back to the referee; perhaps a reader can clarify this.

Anyway, this last exercise provides a radius for our universe that looks surprisingly alike the size of the proton. How does Eq. (6) compare with observations of the proton size? For a comparison with charge-radii observations, a straight average of the radius obtained by various teams (Karshenboim 2000) gives 8.2 (±.3) x $10^{-16}$ m for six observations, of which however there is one as far off as 6.4 x $10^{-16}$ m, while five of them are between 8.09 and 8.90 x $10^{-16}$ m. Two other observations yield 8.05 (±.11) and 8.62 (±.12) x $10^{-16}$ m (Berkeland et al. 1995). It is seen from the high precisions of widely different results that the determination of Eq. (6) could only be for an *equivalent* proton radius, rather than claiming that the proton is a sphere. The word "equivalent" is then for a hypothetical spherical shape of the proton. The proton has for a long time been considered a *fuzzy* sphere having radii between 6 and 10 x $10^{-16}$ m. A better interpretation is with time-dependent shape, perhaps due to internal quark motion (Berkeland et al. 1995).

Another choice instead of H was tried for the computation of Eq. (6), namely the mass of the $^1$H atom, which seems a small increase but the result is grossly off Eq. (6) in view of its precision, at R = 8.192 9019 x $10^{-16}$ m. It is seen that for any value of H larger than that of the proton mass, R would be smaller, and vice versa because of the inverse proportionally in Eq. (2). A smaller size for a larger mass and vice versa? Does Eq. (6) converge on the proton radius as some absolute value?

The equivalent density of the proton, assuming uniformity, follows from Eq. (6) and the proton mass in Table 1,

$$\text{equivalent proton density} = 7.249\ 1169\ (54) \times 10^{17}\ kg\ m^{-3}. \qquad (7)$$

The above radius relation is steep, such that this exercise serves as a confirmation for the mass of our universe. If it would have been for example a factor of 2 larger ($\alpha$ = 2.008), it would have yielded the radius at 1.02 x $10^{-15}$ m, which is out of the question when compared to R and its precision. This result indicates *fine-tuning* for our universe with high resolution compared to nature's large quantization factor of F ~ $10^{19}$ shown between the steps in Table 1 of Eq, (5).

This section brought five topics of support. It shows the linkage of $(hc/G)^\alpha$ with the Planck domain. This yields a theoretical radius of the proton, which is the radius of our universe if it ever would have been at Planck density, and thereby confirms the theory and its finite mass of our universe.

# 5. THE MULTIVERSE

Section 5.1 has observations leading towards the multiverse. The difference with parallel universes is in Sec. 5.2.

## 5.1. Observations

Application of $M(\alpha)$ beyond $\alpha = 2.00$ *must* be explored for four reasons, each sufficient by itself.

• The Supply Problem: where did our universe's observed energy equivalent of $10^{21}$ solar masses come from?

• Another profound mystery with assuming a sole universe is where our physics could have come from. How could something so finely tuned, as we see in stars, have developed in the beginning of our universe when the tools and techniques of evolution were still primitive, so far removed from stars, and all stages lasted exceedingly short times in the expansion? That would have been in contradiction with the characteristics of evolution needing many samples and long times - alike the times and sampling of Darwin's finches (Sec. 6.2).

• Uniformity is observed to third-decimal precision in the 3-K background observations; the temperature of 1.725 K is observed in all directions. How does that uniformity come about? The present model answers this question, as does Inflation theory, but in a simpler manner by the mixing of debris from nearly identical universes.

• Equation (3) yields a mass at any value of exponent alpha because the equation is *open* to all values of alpha.

## 5.2. Difference with parallel universes

Our multiverse is entirely different from parallel universes, for which there are issues of interpretation of quantum mechanics (Penrose in Kruglinski 2009; a detailed discussion of Schrödinger's work is by the French philosopher Bitbol 1995). Furthermore, there is the issue of fine tuning in the theory of stellar structure. Fred Hoyle used to point at the extremely low probability of the fine-tuning for nuclear transitions within stars; the selections and combinations could not have occurred if the physical constants of the elements would have been even slightly different from what they are now. Another case of fine tuning is mentioned for the mass of our universe at the end of Sec. 4. A large literature about many universes then solves the problem statistically by conjecturing that ours happens to have the fine-tuning; these are the many-worlds and parallel universes (Everett 1971, Valenkin 2006).

In contrast, the present paper demonstrates that the fine tuning is done in the *multiverse*, and that by letting Eq. (3) rule without anthropic precepts one finds its universes having their numbers, masses, and physics. Observations do not affect the *macro* world, and the argument that we "better get used to results that seem strange" appears to violate the beauty of truth, aesthetics, and motivations in science (Chandrasekhar 1987; Penrose in Kruglinski 2009).

# 6. THE HISTORY OF OUR UNIVERSE

Section 6.1 details our universe's debris feeding the multiverse. This opens the study of the multiverse as an evolutionary system (Sec. 6.2). The beginning of our universe is in Sec. 6.3.

## 6.1. Decay of our Universe

Everything in our universe ages and decays. Even the proton may have a limited life time, while it is the basic component of every atomic nucleus. Andrei Sakharov (1965) computed its half-life at $10^{50}$ years or longer. Old cold protons and other particles, or their **sub-atomic particles** are part of the universe's decay debris, as are whole galaxies, clusters of galaxies, and whatever other debris such as of old and remnant stars. Dark matter and dark energy must be included as they occur in our universe (Sec. 7.2).

**Old cold photons** are part of the debris. They emerge from sources of radiation; their aging is in terms of expanding out into space to near 0 K. COBE and other surveys observed them at 2.725 K on the way out as a verification of "old cold photons". Within a multiverse, they are conserved. Photons are not particles but waveforms of radiation (Lamb 1995), which may facilitate the physical interpretations, as does the consideration of particles as waveforms (Sec. 7.2).

It is noted how essential for this paper the discovery is of the **acceleration of expansion** at about $5 \times 10^9$ years ago by the teams of Riess et al. (1998) and Perlmutter et al. (1999). If they had found the deceleration that they expected to find, the universe would gravitationally collapse upon itself and the future of our universe would be different.

Information from the interstellar medium (ISM) is useful because some of the same processes are bound to happen in the inter-universal medium (IUM), scaled of course over cosmological scales of space and time. The debris does not have far to go to get into the IUM, because our universe is within it.

The physics of our universe is in unified operation of the *cosmos*' quantum, relativity, gravity, and atomic physics; our four separate theories are still being improved. If our universe resulted from and is decaying back into the inter-universal medium, the IUM must have that physics. The IUM has uniformity through mixing of debris from a large number of universes, such that all universes surviving from that medium have that same physics. The universes also have near-critical mass or they cannot survive the evolution, as they would either collapse (too heavy) or expand rapidly into nonexistence (too light).

The supply to the IUM is continuous, but its composition is now totally different, namely of the above decayed inert debris instead of atomic and molecular active ISM material. It is uniformly mixed because it is fed by debris from many universes [that are all the same according to Eq. (3)], but its space density will again be locally uneven with huge clouds accreting. Nothing stands still in the cosmos - the clouds continue to grow by sweeping the material up during their motion through space. Eventually, self-gravitation will become active, speeding the contraction of the cloud by its increasing gravitational cross-section.

This is *energy-seeking* material, such that the growing proto-universe does not become as hot as a proto-star would have, because the gravitational energy of the compaction re-energizes photons, and re-energizes and/or re-constitutes the atomic components into regular protons, neutrons, and other completed particles. Dark matter and energy must occur in the debris as well.

**6.2. Evolution in the Multiverse**

The multiverse satisfies the four points of Sec. 5.1, and it is an *evolutionary system*. The striking characteristic is the enormously large number of universes, $10^{19}$ at $\alpha = 2.5$, $10^{39}$ at $\alpha = 3.0$, etc. serving to solve for a large number of physical parameters, such as of laws, forces, and the earliest basic particles. Point 18 of Sec. 7.3 has the derivation of $\sim 10^{30}$ years' time scale.

A remarkable characteristic of evolution is the trend towards greater complexity, which usually brings greater capability, so that evolution itself evolves and accelerates. This implies that the multiverse has a beginning, namely of the simplest stage of evolution; this is contrary to no-beginning and no-ending in Gehrels (2007c). The evolution may have an ending as well, at the most advanced stage of evolution, which brings an even grander scale of other *multiverses* fed by the debris of old ones, as Eq. (3) brings with its openness a hierarchy of larger values for alpha.

For a detailed study, observations of evolution in the inorganic domain are in a text that shows 13 of them (Gehrels 2007c). There seems to be an overall trend towards greater complexity, which usually brings greater capability. Inorganic and organic evolutions show the same evolving evolution with tools of increasing sophistication such as reproduction in the organic. Nature appears to be making trials towards continuation, for what can proceed - will it yield survival, or is the trial in error? The trials depend on the environment - it is *natural selection* - with random, slightly different features emerging over long time scales.

One can confine the modeling in the multiverse to a relatively small volume, but still with a multitude of universes, and considering this as a closed system in which everything is conserved; an example may be the Local Group of universes with $3 \times 10^{19}$ universes at $\alpha = 2.50$ in Eq. (3). Many physical parameters may be established in trial-and-error evolution, such as the four atomic forces and basic particles, because there have been many universes functioning as species and there are long periods of time for trials small changes that may occur ($\sim 10^{30}$ years, Sec. 7.3). Any failures vanish back into the IUM when a universe happens to originate with characteristics that error too much. The evolution has all the universes originating at the present time with the characteristics of Sec. 5.1, but slowly over time scales of $\sim 10^{30}$ years there might be some change. The mixing of debris in the IUM from a large number of universes causes the fact that all new universes of a certain epoch have the same physics, which might be evolving, but exceedingly slowly.

**6.3. Beginning of Our Universe**

The growth of the IUM's clouds, at the end of Sec. 6.1, comes to a halt in a remarkable way. Imagine a gravitationally layered spherical cloud of about the size of the Mars orbit, and with uniformly distributed debris; the compressed clusters of galaxies may be recognizable because of their large masses. When the *central* region comes to the density of $10^{18}$ kg m$^{-3}$, the old cold photons re-energize first, because this is the density at which photons formed at age $t \sim 10^{-6}$ s in standard modeling. The briefly emerging burst of photons apparently was the recent WMAP discovery of a feature of wider curvature than that of the 3-K radiation (Hinshaw 2009).

However, that central region also had sub-atomic particles, which may have taken a little longer to be re-constituted as well as re-energized; $10^{18}$ kg m$^{-3}$ is the density of protons [Eq. (7)] and $t = 10^{-6}$ s is the epoch of their birth in standard modeling. Dark energy, and perhaps dark mass, may have acted as a catalyst in this intricate process (Sec. 7.2).

In any case, as soon as the protons were formed, they retarded the photons from getting out through high opacities due to their multiple scattering, also by neutrons, etc. The scattering continued until age 380,000 when at space density of $\sim 10^{-19}$ kg m$^{-3}$ the electrons and protons re-combined to make atoms again, having enormous internal space to let the photons through.

There are other observations related to this beginning of our universe:
• The history depends on the assumption that the energy-seeking feature of the debris particles is effective enough to keep the temperature low enough for their old previous characteristics to survive, such as the galaxy clustering. The situation needs to be modeled in detail for temperature, with dark matter and varying presence of dark energy (Sec. 7.2).
• There is the peculiar fact that only 4.6% of the mass of our universe is baryonic, visible matter. Why it is so low has been a persistent query, but now it may be seen from the simple spherical geometry of the cloud; the central volume at $10^{18}$ kg m$^{-3}$ is small in a gravitational globe, apparently limited to allow only 4.6% of the mass to be re-energized. I will call it the "4.6% volume", but the dark mass and dark energy are present. Outside of the 4.6% volume, the re-energizing and re-constitution cannot take place because of insufficient density (Sec. 7.2).
• The timing of $t \sim 10^{-6}$ s comes from the standard models for the photon as well as proton generation.
• The fact that there are no observations beforehand, but several afterwards, is also consistent with this timing.
• This establishes the beginning of our baryon universe, at $\sim 10^{-6}$ secs later, *i.e.* $\sim 10^{37}$ Planck times later than would have occurred in Big-Bang theories. The earlier atomic physics predicted in these theories occurred *in the multiverse*.

# 7. CONCLUSIONS AND FUTURE WORK

Section 7.1 has Schwarzschild's limitation of how close to t = 0 our universe could have existed. It is t ~ $10^{-6}$ s later, a surprising conclusion confirmed in Sec. 7.2. The paper ends with a summary of 37 conclusions or suggestions for future work (Sec. 7.3).

**7.1. Schwarzschild's Proof**

Schwarzschild's limitation shows how far we may theorize back in time for modeling towards the Big Bang, until the mass concentration would produce a black hole. Putting that in the proper order: our universe cannot have existed before that time. It gives an upper limit of a mass and radius combination below which radiation cannot escape because the object is a black hole. Karl Schwarzschild wrote it in 1916 as a part of his detailed derivation of one of Einstein's equations. The arithmetic is simple because for light to escape, its kinetic energy must be greater than the local gravitational potential. For the velocity of light, c, it follows that the limiting radius is $R_S = 2GM/c^2 = 3 \times 10^8$ ly. Table 2 shows the comparison of $R_S$ with radius $R = (3M/4\pi\sigma)^{1/3}$ for a uniform sphere with density σ; M is $1.13 \times 10^{78}$ proton masses.

The first line of the Table is for the epoch when the universe's radiation did indeed escape, at age 380,000, $R/R_S = 1$. Schwarzschild's limit does not seem precise because standard theories predict the density to be $10^{-19}$ kg m$^{-3}$ at that time, not $10^{-23}$. However, the precision of these predictions is low, and the effect is small because if $10^{-19}$ were used in the calibration of $R_S$, the following ratios are still $10^{-14}$, and $10^{-38}$ instead of $10^{-40}$.

**Table 2. Universe Radii and Schwarzschild Radii**

| $R/R_S$ | t | σ | R(ly) |
|---|---|---|---|
| 1 | 380,000 y | $10^{-23}$ | $10^8$ |
| $10^{-14}$ | $10^{-6}$ s | $10^{18}$ | $10^{-5}$ |
| $10^{-40}$ | 0 | $10^{96}$ | $10^{-31}$ |

The second line is for the above t ~ $10^{-6}$ s, using proton density of $10^{18}$ kg m$^{-3}$ to derive R [Eq. (7)]. Because $R/R_S = 10^{-14}$ is so very negative, the Schwarzschild radius indicates a black hole, but our universe is *not* a black hole. Furthermore, for photons to escape at age 380,000 as is well established, the generation near the center *must have occurred much earlier*. In the case of the sun, it takes a million years for a photon generated at its center. Now, the shorter time of 380,000 years may be right, even though the body is much larger than the sun, because the medium is *expanding*. The photons' scattered journey was rapidly speeded up because it went through diminishing density, eventually as low as the above ~$10^{-19}$ kg m$^{-3}$, which by then was the density of the whole cloud. What had caused the enormous effect of pushing the whole cloud apart, *i.e.* what caused the expansion of intergalactic space? The latter has been the classical question ever since the 1920s, and the next Sec. 7.2 has the answer.

In the third line, the Planck density of Eq. (4) is used to derive R and thereby $R/R_S$ (as if our universe were ever at t = 0). Because $R/R_S = 10^{-40}$, the Schwarzschild limit calls resoundingly for a black hole. However, our universe is not a black hole and this time the controversy is not resolvable; the standard models hypothesize only *extremely basic* particles for the earliest times between t ~ $10^{-43}$ and $10^{-7}$ s, no photons.

Before t ~ $10^{-6}$ s, equal to t = 0 on the new clock, the physical evolution occurred in the multiverse, and after t ~ $10^{-6}$ s our understanding of it is back on track of the standard models for our universe and for particle physics. Our universe began with photons, protons, neutrons, etc., ready to go on their observed paths.

## 7.2. Observations of Dark Energy

The greatest clue we have left over is that the observations of dark energy terminated as if it been used up in making our universe. It is seen from the comparison of the next two paragraphs; the percentages are in terms of mass (Hinshaw 2009).
1. At the present time, the baryons amount to only 4.6%. Neutrinos have less than 1%, while 23% is not-observable dark matter. The dominant 72% is in some form of dark energy.
2. When the universe had age 380,000, our universe amounted to 12% atoms, 15% photons, and 10% neutrinos. Not observable but otherwise derived to be present was 63% dark matter, and there was little or no observation of dark energy.

The difference of the dark-energy numbers is now used to answer the classical question, "What caused the expansion of our universe?" It is noted that *causing* the expansion requires the same effect and therefore probably the same physical action as *accelerating* the expansion, and the latter is in the literature as having been caused by dark energy. The conclusion is then that the expansion is caused by dark energy.

The "4.6% volume" of Sec. 6.3 was small, but its original burst of radiation may have stopped the accretion and perhaps it began to reverse the in-fall into an expansion by its radiation pressure. The dark energy must have participated as it was present there with its (accelerating of) expansion action. By the dominance of 72% over 6.4%, and by its general presence, the dark energy was dominant in bringing the expansion of the entire universe about. The observation on which this conclusion is based is the difference between 380,000 and a million years in the discussion of Sec. 7.1. Without expansion, that interval would have been much larger than a million years for the photons to escape, because the mass of our universe is so much larger than that of the sun, $10^{21}$ times larger. The radius of our universe at 380,000 years was $10^{15}$ times larger than that of the sun.

## 7.3. Summary Listing

During the years of searching for this history of our universe, new insights and observations have invariably brought progress, and this process has not stopped as yet; new ideas keep coming. This seems an indication of truth for the model, as are its common sense, internal consistency, and beauty (Chandrasekhar 1987). If so, one can turn the reasoning around, assuming this history is nearly correct and thereby making predictions for new observations and analysis. This is more difficult and it depends on the researcher and facilities, but some of the following conclusions may be helpful. Some points do not occur in the above text; this listing assembles material for new physics.

1. The $M(\alpha) = (hc/G)^{\alpha}$ equation in proton masses H, has that proton mass at its foundation, it is connected to the Planck domain and is summarized as a *universal* Planck mass; it compares well with observations for primordial stars and universe.
2. It uses h, c, G, and H derived by Chandra from physical laws representing aspects of quantum, relativity, gravity, and atomic physics together in a unified manner.
3. The application is without any interpretation of quantum mechanics.
4. The multiverse is a quantized hierarchy of exponentially increasing numbers of universes.
5. There is a quantization factor between masses, $F = 3.3 \times 10^{19}$, which is the same between the Planck length and the proton radius of Eq. (6).
6. Study of the basics of quantization are encouraged by the $M(\alpha)$ findings.
7. Further study is needed whether or not planetesimals, and perhaps galaxies (Gehrels 200b), should be included in the discussion of $M(\alpha)$.
8. The new definition of the Planck mass is its role in the mass scaling of the cosmos.
9. Observations of the proton size were verified, and the procedure confirms that $\alpha = 2.00$ in $(hc/G)^{\alpha}$ gives the finite mass of our universe as $1.131\ 79(35) \times 10^{78}$ proton masses = $9.5172 \times 10^{20}$ solar masses.

10. The equivalent spherical radius of the proton is 8.197 3725(20) x $10^{-16}$ m.
11. What causes the time-dependence of the shape of the proton?
12. The treatment of this paper indicates that the Planck constant is h, not $\hbar = h/2\pi$, or the equivalent radius of the proton would be 2.4 x $10^{-16}$ m, an O-type star would have 2 solar masses, and both are out of the question. It might be appropriate to call $\hbar$ the Dicke constant, or a mathematical convenience.
13. It also indicates that the cosmological constants h, c, G, and H are constant over a considerable range of location and ~$10^{30}$ years. The universe is however evolving such that the constants will also change albeit imperceptibly slowly (Sec.7.2).
14. Five reasons show why there is a multiverse with specific mass and physics for all universes.
15. It follows that all universes have the same h, c, G, H physics, except for small differences depending on when they emerged from the evolving multiverse.
16. All universes must have near-critical mass to survive.
17. The scale of numbers in the inter-universal medium is larger than in the interstellar medium by a factor of ~$10^{19}$, which is the quantization factor F of Eq. (5). A coarse estimate of the cosmological *time* scale follows from Sakharov's >$10^{50}$ years as an upper limit. The lower limit is observed for stars on the order of $10^{11}$ y for the slowest. The time scale within the multiverse may therefore be somewhere in between the two limits, very roughly at ~$10^{30}$ y, with a factor $10^{30}/10^{11}$ ~ $10^{19}$ as in Eq. (5) again.
18. The basic paper by Sakharov may need to be re-visited.
19. The cosmological foundation of our world and its physics is in trial-and-error evolution within its hierarchy of universes.
20. The M($\alpha$) theory includes that of the origin of our physics because the inter-universal medium must have it, and therefore all universes as well.
21. Fred Hoyle's fine-tuning of the nuclear transitions within stars follows because the continuing trial-and-error evolution within the multiverse produces finely tuned universes to begin with.
22. Universes apparently decay into sub-atomic particles and cold photons over cosmological time scales and at near-absolute-zero temperatures.
23. The accelerated expansion of intergalactic space brings mixing of universes over long cosmological times.
24. The inter-universal medium thereby consists of all possible components, including galaxies that are gravitationally held together, and clusters of galaxies. Protons, neutrons, electrons, dark matter, and dark energy are included as well as stellar remnants.
25. The question arises if we might not see in our present universe some of the debris of other old universes. Basu (2009) has been inspecting the literature for decades for QSOs, AGN, high and very high redshift galaxies, BL Lac objects, host galaxies of GRBs and SNe Ia that have blue-shifted spectra and has published ~200 blueshifts.
26. It appears necessary to model the accretion of our proto-universal cloud and the beginning of our universe, particularly to see if characteristics melt away; dark energy dark matter should be included.
27. The wave interpretation of Schrödinger, Lamb, and others appears important for such studies.
28. The Schwarzschild radius appears to prohibit any stages earlier than t ~ $10^{-6}$ s.
29. The classical wondering why the baryons have only 4.6% of the total composition of our universe follows readily from the spherical geometry of the proto-universe.
30. The re-configuration into photons may have caused a "Photon Burst" beginning of our universe near $10^{17}$ kg $m^{-3}$ [Eq. (6)] and t ~ $10^{-6}$ s (on the clock of the old standard model). WMAP may have discovered it as the radiation signature with a wider curvature than that of the 3-K radiation.
31. Because the universes begin at ~ $10^{-6}$ secs instead of t = 0, the Big Bang, Inflation, Strings,

and the earliest atomic theories are no longer needed.

32. It appears likely that the dark energy, in addition to causing the acceleration of expansion, caused the expansion to begin with.

33. It appears likely that the nature of dark energy can be derived from the above.

34. It appears possible that the nature of dark matter can be derived from the above.

35. The present model needs to be compared to the recent results obtained with WMAP and others, especially where agreement with inflation theory has been found. The expectation is that the same agreement will be found with the model of this paper.

36. This paper uses 25 observations, some of them being sets of observations.

37. New disciplines may appear in pursuing predictions and suggestions for future work.

## ACKNOWLEDGMENTS

Over the eight years of development there have been hundreds of e-mails and discussions. I thank so many colleagues and referees, my hosts such as Leonid Ksanfomality, Neil Gehrels for providing information, and Kees de Jager for urging me to use the Schwarzschild radius.